\documentclass[a4paper,pre,superscriptaddress,longbibiliography]{revtex4-2}

\usepackage{graphicx}
\usepackage{comment}
\usepackage{amsmath,amssymb,amsthm} 
\usepackage{verbatim}
\usepackage{float}

\begin{document}

\title{Even strong energy polydispersity does not affect the average structure and dynamics of simple liquids}

\author{Trond S. Ingebrigtsen}
\email{trondingebrigtsen@hotmail.com}
\author{Jeppe C. Dyre}
\email{dyre@ruc.dk}

\affiliation{Glass and Time, IMFUFA, Department of Science and Environment, Roskilde University, Postbox 260, DK-4000 Roskilde, Denmark}\email{trond@ruc.dk}

\begin{abstract}
  Size-polydisperse liquids have become standard models for avoiding crystallization, thereby enabling studies of supercooled liquids and glasses formed, e.g., by colloidal systems. Purely \textit{energy} polydisperse liquids have been studied much less, but provide an interesting alternative. We here study numerically the difference in structure and dynamics obtained by introducing these two kinds of polydispersity into systems of particles interacting via the Lennard-Jones and EXP pair potentials. To a very good approximation, the average pair structure and dynamics are unchanged even for strong energy polydispersity, while this is not the case for size-polydisperse systems. When the system at extreme energy polydispersity undergoes a continuous phase separation into lower and higher particle-energy regions whose structure and dynamics are different from the average, the average structure and dynamics are still virtually the same as for the monodisperse system. Our findings are consistent with the fact that the distribution of forces on the individual particles do not change when energy polydispersity is introduced, while they do change in the case of size polydispersity. A theoretical explanation of our findings remains to be found, however.\\
\end{abstract} 

\maketitle
\newpage

\section{Introduction}

Polydispersity is often introduced in models of supercooled liquids and glasses as a means to avoid crystallization and hence facilitate formation of the glass phase \cite{bagchi2012,wolynes2012,dickinson1978,blum1979,salacuse1982,gualtieri1982,ginoza1997,evans1999,sollich2002,fasolo2003,jacobs2013,frenkel1986,kofke1986,kofke1987,stapleton1990,auer2001,kristof2001,murarka2003,wilding2005,wilding2005b,wilding2006,kawasaki2007,abraham2008,wu2012,jacobs2013,sarkar2013,ogarko2013,will2013,ashton2013,nguyen2014,phillips2014,sarkar2014,koningsveld1971,cowell1982,weeks2000,ye2005,watanabe2008,ballesta2008,banerjee2012,sacanna2013,palberg2014}. In colloidal fluids size polydispersity has been utilized to destabilize the crystal phase \cite{frenkel1986}, and for obtaining optimal metallic glass formers four or five different components are often introduced \cite{wei2019}. As other examples, Debenedetti and coworkers studied polydisperse systems in connection with determining the connectivity, volume, and surface area of the void space of sphere packings \cite{sas97,sas98}, as well as the formation of microspheres following rapid expansions of supercritical solutions \cite{tom91}. Recently, polydisperse systems have come to play a prominent role in glass science because they can be equilibrated by the swap algorithm at much lower temperatures than is attainable by standard Molecular Dynamics simulations \cite{nin17}.

Size polydispersity has mainly been studied, but recently purely energy-polydisperse systems have become the target of investigations as they pose interesting model systems \cite{ingebrigtsen2016}. Previous studies of ``all particles are different'' models \cite{shagolsem2015a,shagolsem2015b,rabin2016} found that energy polydisperse systems self-organize according to particle energies, an effect that becomes more pronounced with decreasing temperature. References \onlinecite{shagolsem2015a} and \cite{shagolsem2015b} studied two-dimensional energy-polydisperse Lennard-Jones (LJ) systems and found very small, yet clearly discernible differences between average properties such as melting curves, etc., compared to those of the one-component LJ fluid with the same average interaction energy. Polydispersity has also been studied \cite{ingebrigtsen2015,ingebrigtsen2016,ing21} with respect to their isomorphs, which are curves in the thermodynamic phase diagram along which structure and dynamics are invariant to a good approximation \cite{IV,dyr18}.

In this paper we compare size and energy polydispersity with respect how they affect the structure and dynamics of simple liquids, primarily Lennard-Jones systems. In contrast to the well-known significant effects of size polydispersity, we find that the average structure and dynamics are largely unchanged when introducing energy polydispersity, a result that applies even when the system phase separates. We argue that this finding is related to the fact that the average force distribution is virtually unchanged by the introduction of energy polydispersity, which is not the case for size polydispersity.

The next section provides details on the models studied and the simulations used, after which the size vs. energy polydispersity comparison is presented.

\section{Simulations and models}

We studied size and energy polydisperse liquids in the \textit{NVT} ensemble using the RUMD package \cite{rumd} for efficient GPU-based simulations. In most of the paper the interaction potential between particles $i$ and $j$ is the well-known Lennard-Jones (LJ) pair potential, 

\begin{equation}\label{lj}
  v_{ij}(r) = 4\varepsilon_{ij}\Big[\Big(\frac{\sigma_{i j}}{r}\Big)^{12} - \Big(\frac{\sigma_{i j}}{r}\Big)^{6}\Big]
\end{equation}
in which $\sigma_{i j}$ and $\varepsilon_{i j}$ are a length and an energy parameter, respectively. We shall (mostly) assume the standard Lorentz-Berthelot mixing rule\cite{tildesley}, i.e., that

\begin{align}\label{mix}
  \sigma_{i j} & = \frac{\sigma_{i} + \sigma_{j}}{2},\\
  \varepsilon_{i j} & = \sqrt{\varepsilon_{i} \varepsilon_{j}},
\end{align}
where $\sigma_{i}$ and $\varepsilon_{i}$ are the length and energy parameters associated with particle $i$. Unless nothing else is stated, the flat (box) distribution is assumed for both size and energy polydispersity. 

$N$ = 32,000 particles were simulated. Following the convention in the field, for the probability distribution $p(x)$ the polydispersity $\delta_x$ is defined\cite{ingebrigtsen2016} by 
$\delta_x^2=(\langle x^2\rangle-\langle x\rangle^2)/\langle x\rangle^2$. In words, $\delta_x$ is the ratio between the standard deviation and the average. Throughout the paper we use the unit system in which $\langle \varepsilon \rangle$ = 1 and $m$ = 1. In these units the time step is given by $\Delta t$ = 0.0025. We employed a cut-and-shifted potential with cut-off at $r_{\rm c}$ = 2.5$\sigma_{ij}$ for the $ij$ particle interaction. 

The changes in structure and dynamics introduced by polydispersity are probed via the average radial distribution function (RDF) and the average self-part of the intermediate scattering function (ISF).

\section{Average structure and dynamics of polydisperse LJ systems}

\begin{figure}[H]
  \begin{minipage}[t]{0.45\linewidth}
    \centering
    \includegraphics[width=68mm]{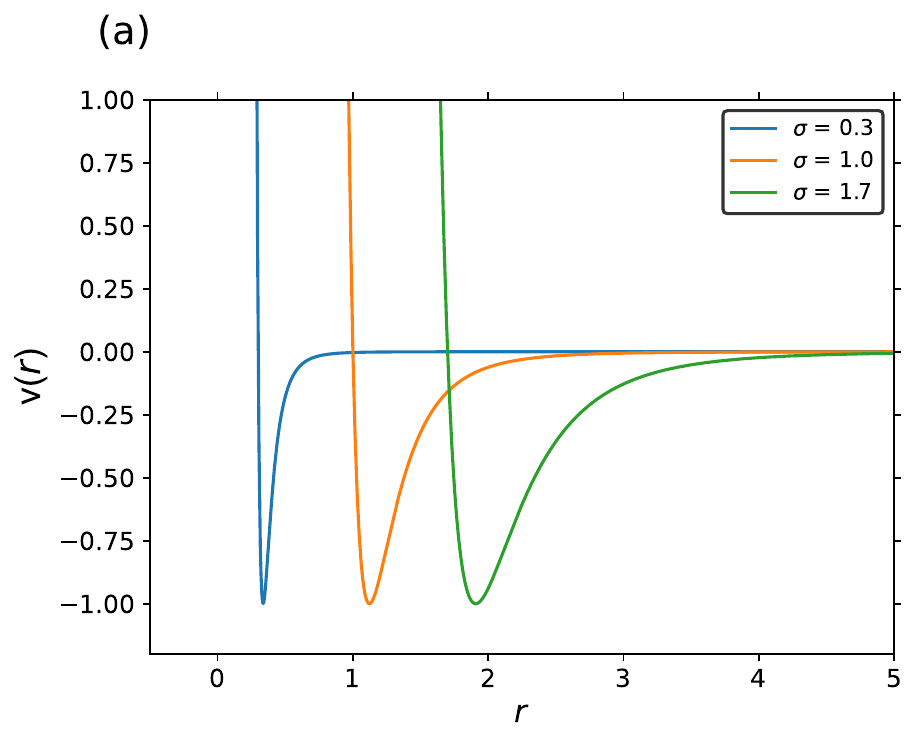}
  \end{minipage}
  \begin{minipage}[t]{0.45\linewidth}
    \centering
    \includegraphics[width=68mm]{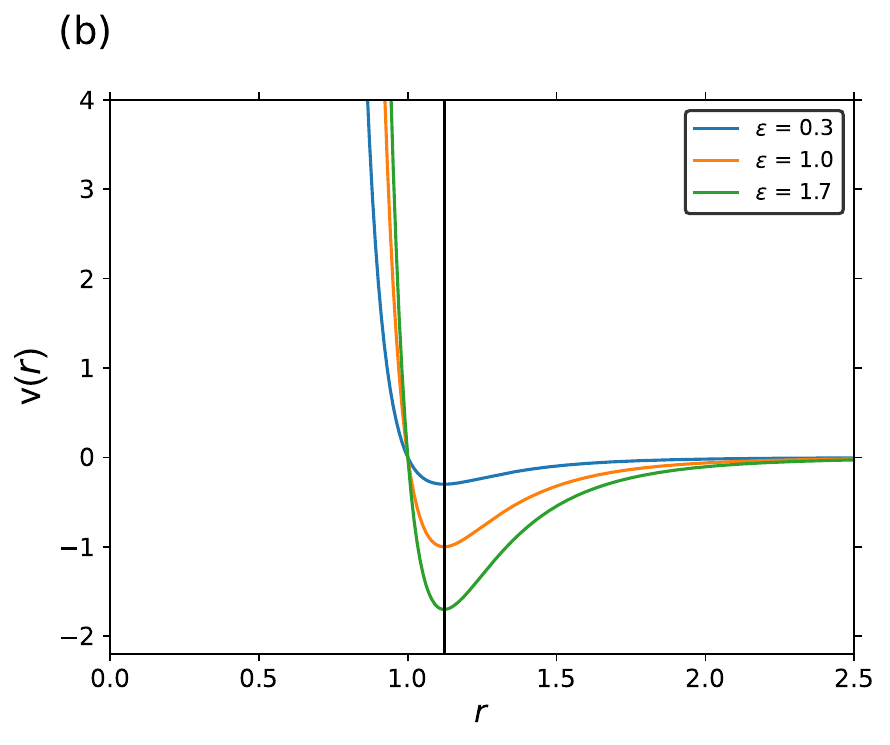}
  \end{minipage}
  \caption{Size and energy polydisperse LJ potentials. (a) LJ pair potentials at 40\% box distribution size polydispersity, showing the two extremes as blue and green, respectively, as well as an in-between case (orange). (b) LJ pair potentials at 40\% enegy polydispersity, showing again the two extremes and an in-between case. Note that in this case, all three pair potentials have minimum at the same pair distance (marked by the vertical line at $r=2^{1/6}\sigma$).}
  \label{potential}
\end{figure}

Figure \ref{potential} shows the LJ pair potentials for the highest size and energy polydipersities studied, $\delta \cong 40\%$, by giving the extremes of the flat (box) distribution. One notes that the potentials vary significantly in both cases. Compared to most experiments,  $\delta = 40\%$ is a sizable polydispersity that covers a range of sizes/energies of more than a factor of five.

\begin{figure}[H]
  \begin{minipage}[t]{0.45\linewidth}
    \centering
    \includegraphics[width=70mm]{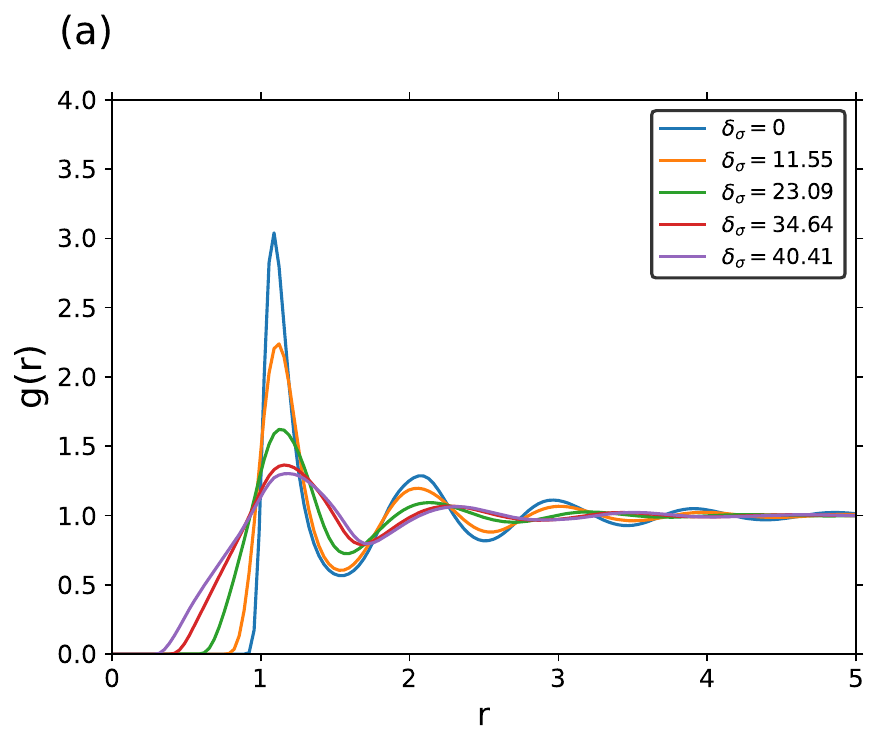}
  \end{minipage}
  \begin{minipage}[t]{0.45\linewidth}
    \centering
    \includegraphics[width=70mm]{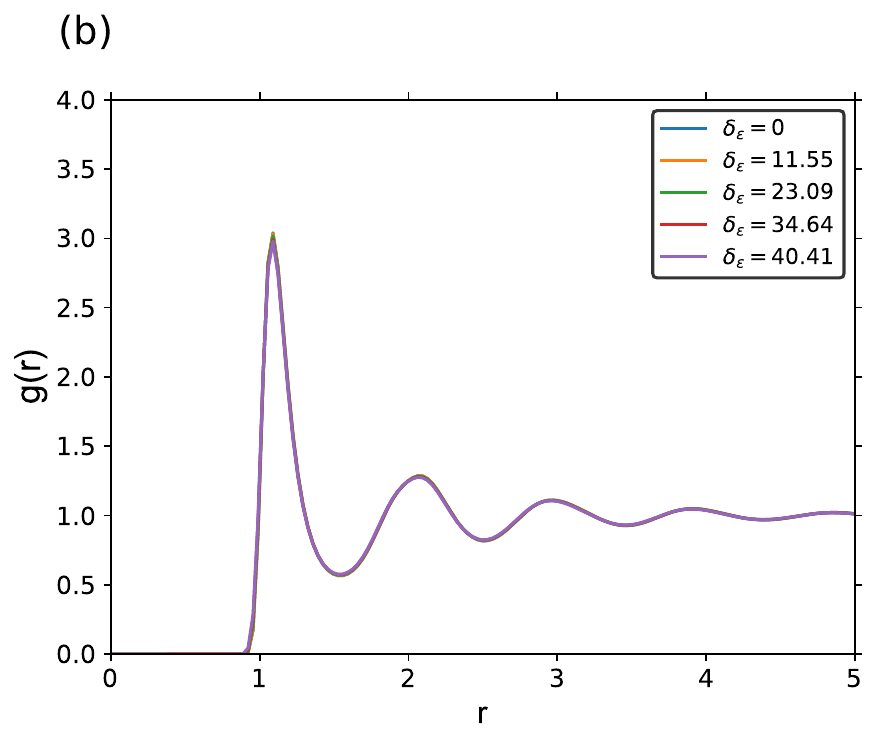}
  \end{minipage}
  \begin{minipage}[t]{0.90\linewidth}
    \centering
    \includegraphics[width=70mm]{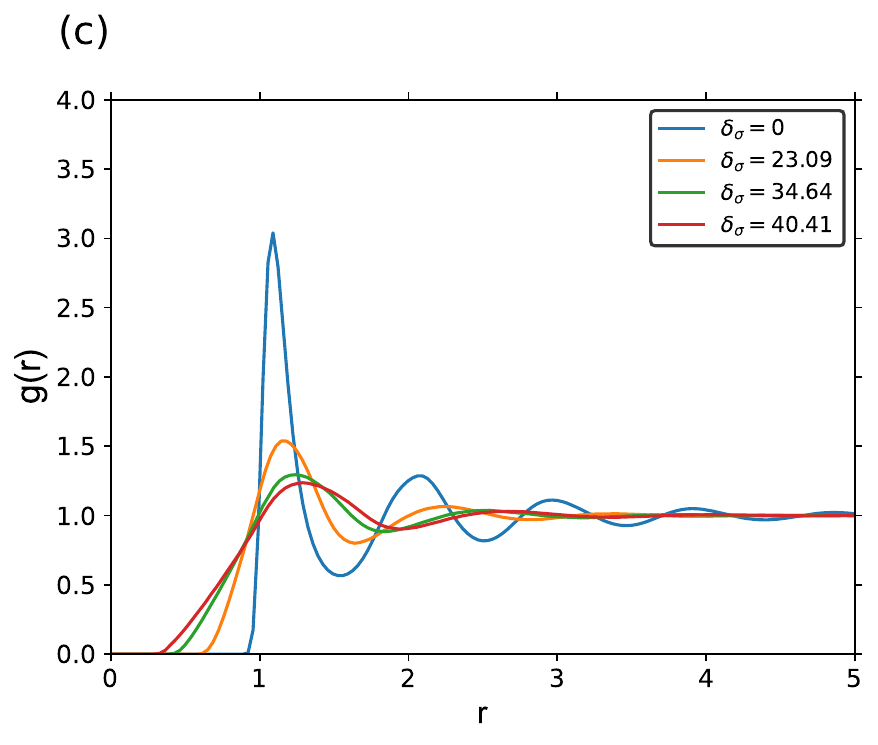}
  \end{minipage}
  \caption{Effect of size and energy polydispersity on the average RDF at the state point $(\rho,T)= (0.85,0.70)$. (a) Size polydispersity. (b) Energy polydispersity; here the average RDF is almost independent of the degree of polydispersity. (c) Effect of size polydispersity at constant packing fraction instead of constant density.}
  \label{rdf}
\end{figure}

Next, we present data for the average pair structure and dynamics at the state point $(\rho,T) = (0.85, 0.70)$ in which $\rho$ is the particle density and $T$ the temperature. For the single-component LJ system this is a typical liquid state point located near the triple point. Figure \ref{rdf} shows how the average RDF is affected by increasing polydispersity in the two cases; the polydispersity is introduced such that the average $\sigma$ and $\varepsilon$, respectively, equals unity, i.e., the box distributions used are symmetric around unity. Figure \ref{rdf}(a) shows the effect of introducing size polydispersity, which strongly influences the average RDF. In contrast, (b) shows that for energy polydispersity there is almost no change in the average RDF. 

That size polydispersity affects the average RDF (Fig. \ref{rdf}(a)) is not unexpected. In fact, because we keep the particle density constant and vary $\sigma$ according to a box distribution with fixed mean, a larger polydispersity leads to a larger packing fraction. This is not the whole explanation, however, since even when keeping the packing fraction constant, one still observes a strong variation of the average RDF (Fig. \ref{rdf}(c)). 

\begin{figure}[H]
  \begin{minipage}[t]{0.45\linewidth}
    \centering
    \includegraphics[width=70mm]{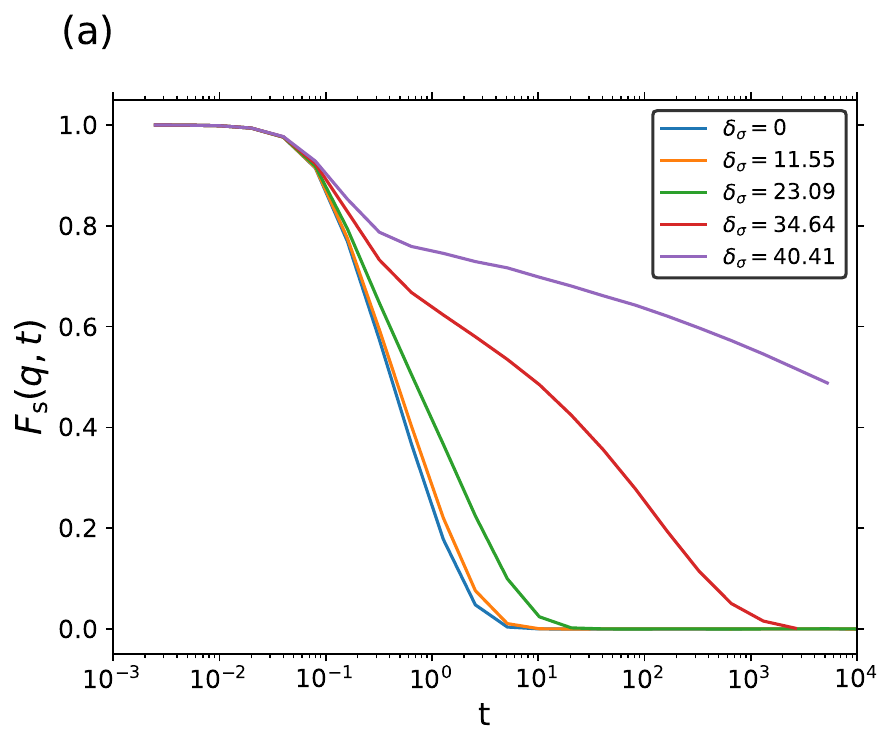}
  \end{minipage}
  \hspace{0.5cm}
  \begin{minipage}[t]{0.45\linewidth}
    \centering
    \includegraphics[width=70mm]{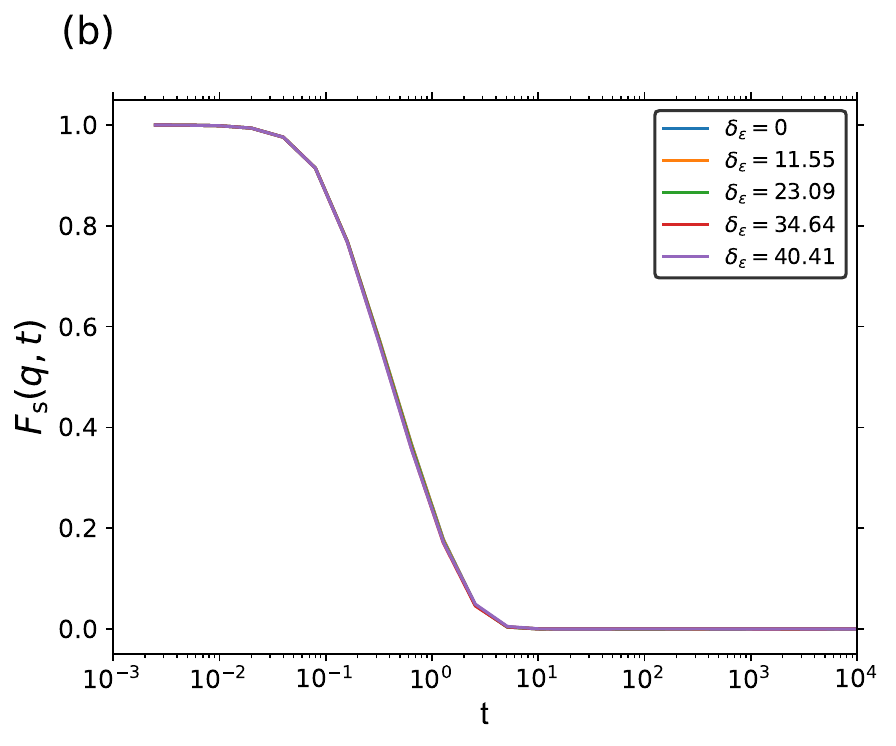}
  \end{minipage}
  \caption{Effect of size and energy polydispersity on the incoherent intermediate scattering function (ISF) $F_s(q,t)$ at the state point $(\rho,T)= (0.85,0.70)$ for the wave vector corresponding to  the first peak of the static structure factor of the monodisperse system. (a) Size polydispersity. (b) Energy polydispersity.}
  \label{intermediate}
\end{figure}

Turning to the dynamics, Fig. \ref{intermediate} presents data for the incoherent intermediate scattering function (ISF), $F_s(q,t)$, in which ``s'' signals the self part, $q$ is the wave vector, and $t$ is the time. The results are similar to those of Fig. \ref{rdf} with a substantial change in the dynamics for size, but not for energy polydispersity. We note a significant slowing down for large size polydispersity, which undoubtedly is caused by the above discussed increased packing fraction.

\begin{figure}[H]
  \begin{minipage}[t]{0.45\linewidth}
    \centering
    \includegraphics[width=70mm]{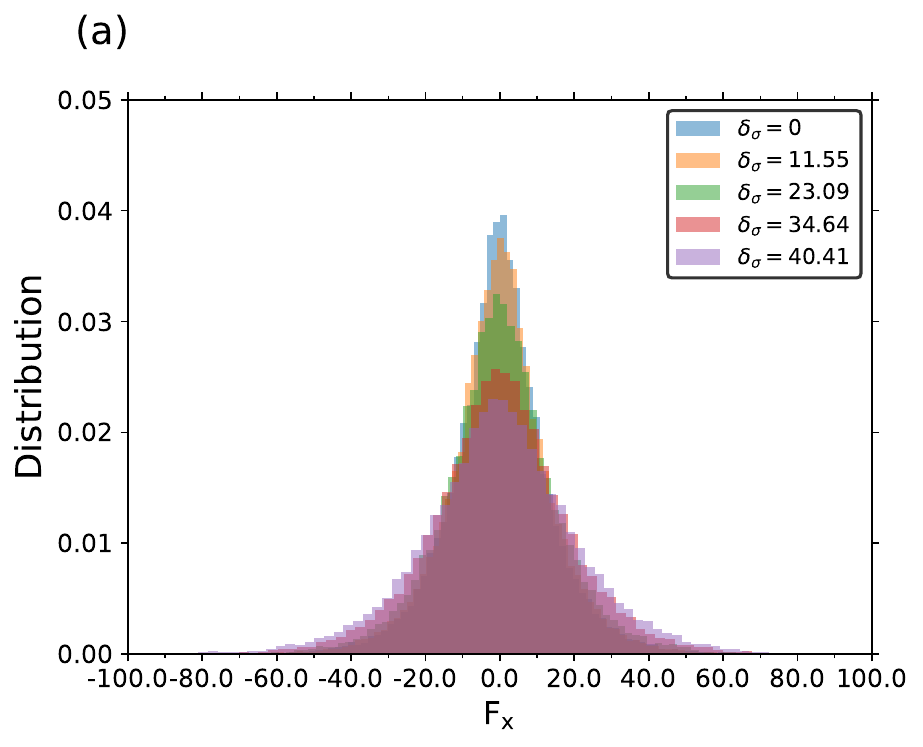}
  \end{minipage}
  \hspace{0.5cm}
  \begin{minipage}[t]{0.45\linewidth}
    \centering
    \includegraphics[width=70mm]{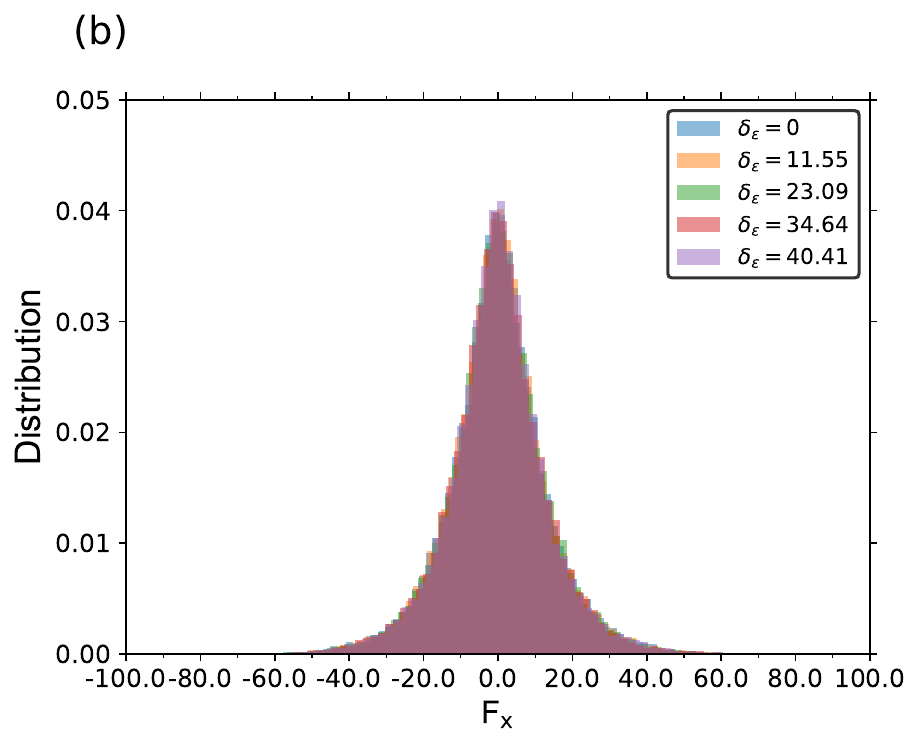}
  \end{minipage}
  \centering
  \begin{minipage}[t]{0.45\linewidth}
    \centering
    \includegraphics[width=70mm]{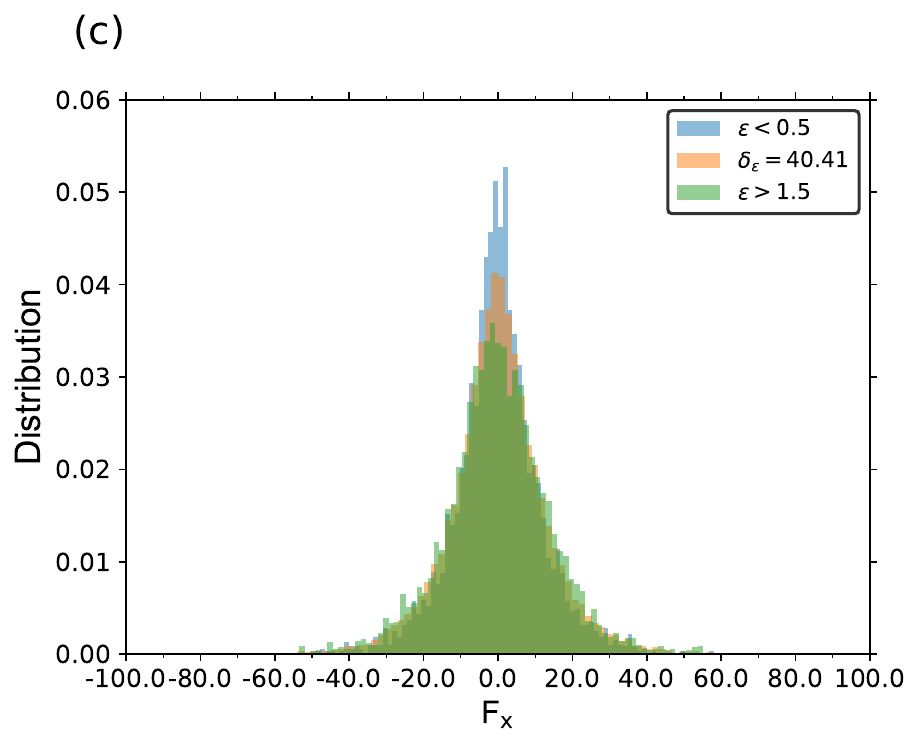}
  \end{minipage}
  \caption{Effect of size and energy polydispersity on the distribution of $x$-components of the particle forces, $F_{x}$. (a) Size polydispersity. (b) Energy polydispersity. (c) Energy-resolved force distribution for the highest energy polydispersity (40\%). In (c) the higher-energy particles have a broader force distribution than the lower-energy particles. The barely visible orange histogram is the average force distribution. These normalized probability distributions are all well fitted by a Gaussian, meaning that the width, which in some cases cannot be visually identified, is inversely proportional to the height.}
   \label{force}
\end{figure}

Figure \ref{force} displays the distribution of the $x$-components of the particle forces. The distribution is visibly affected by size polydispersity ((a)), but little by energy polydispersity ((b)). This suggests that not only is the average pair structure and dynamics invariant, so are also other \emph{average} structural and dynamical quantities because one expects the average force to determine these quantities.

Probing the energy-resolved force distribution in Fig. \ref{force}(c) for the case of energy polydispersity, we observe a dependence on the identity of the particles. That is not surprising because one expects larger forces on particles of larger energies $\varepsilon_i$, which is precisely what is seen, but it only serves to emphasize the mystery of why the average force is insensitive to the introduction of energy polydispersity.

\section{Results for other state points, mixing rules, distributions, pair potentials, and for a binary mixture}

This section investigates the generality of the above findings for the LJ pair potential.

\subsection{Results at three other state points}

\begin{figure}[H]
  \begin{minipage}[t]{0.45\linewidth}
    \centering
    \includegraphics[width=70mm]{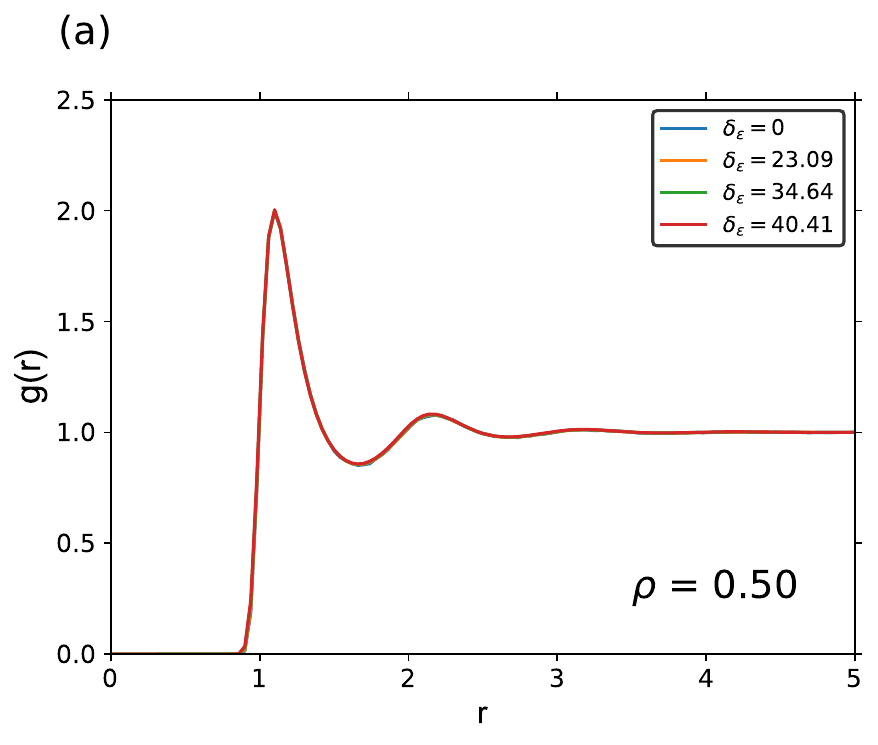}
  \end{minipage}
  \hspace{0.5cm}
  \begin{minipage}[t]{0.45\linewidth}
    \centering
    \includegraphics[width=70mm]{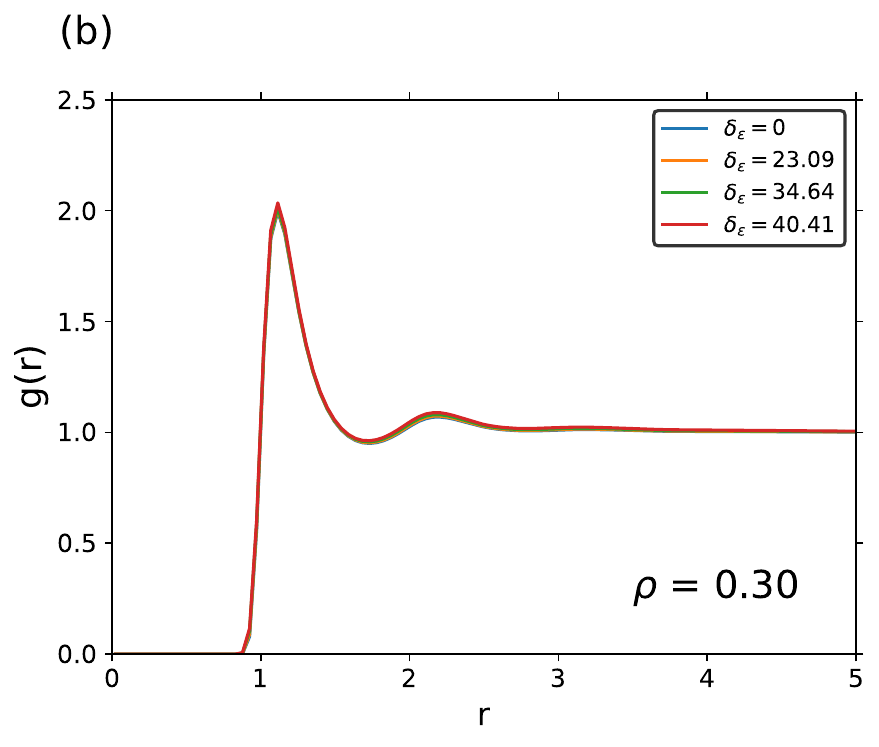}
  \end{minipage}
    \centering
  \begin{minipage}[t]{0.45\linewidth}
  \centering
    \includegraphics[width=70mm]{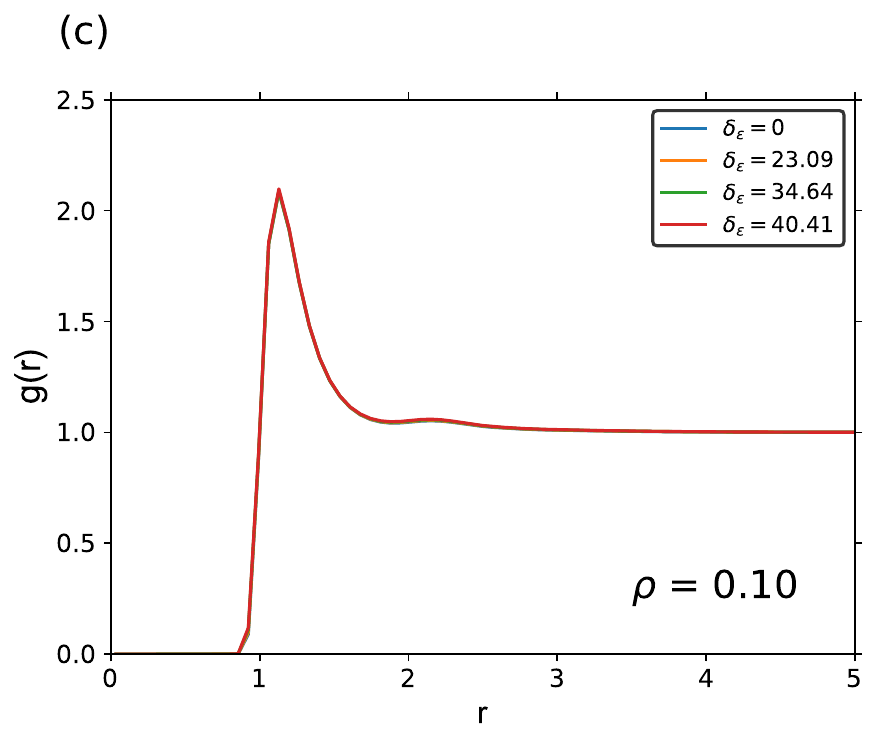}
    \end{minipage}
  \caption{Effect on the average RDF of different energy polydispersities at $T$ = 1.3 at three different densities. (a) $\rho$ = 0.50. (b) $\rho$ = 0.30, which is close to the critical point of the monodisperse LJ system. (c) $\rho$ = 0.10. In all cases the energy polydisperse systems have an average RDF virtually identical to that of the monodisperse system.}
  \label{lowerdensity}
\end{figure}

Is the finding of little effect of energy polydispersity particular to the state point studied? Figure \ref{lowerdensity} displays the average RDFs for the densities $\rho$ = 0.10, 0.30, 0.50 at $T$ = 1.3. We find little effect of introducing energy polydispersity, whereas size polydispersity has a huge effect on the average RDFs at all three state points (results not shown). Note that the $\rho=0.30$ state point is just above the critical point for the monodisperse system; while this proximity to a second-order phase transition weakens the collapse, the effect is minor.

\begin{figure}[H]
  \begin{minipage}[t]{0.45\linewidth}
    \centering
    \includegraphics[width=70mm]{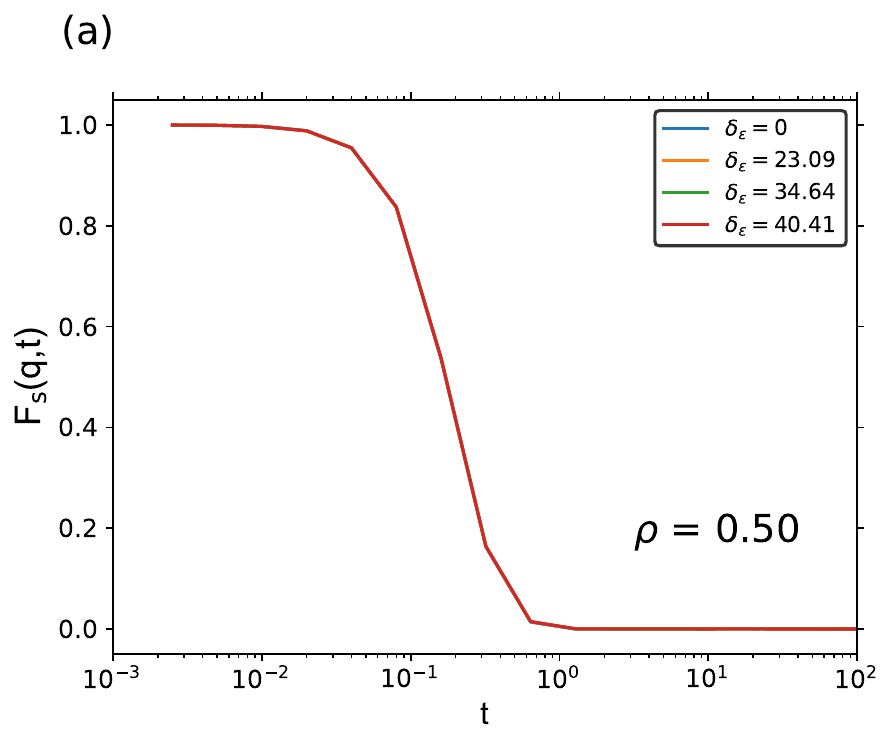}
  \end{minipage}
  \hspace{0.5cm}
  \begin{minipage}[t]{0.45\linewidth}
    \centering
    \includegraphics[width=70mm]{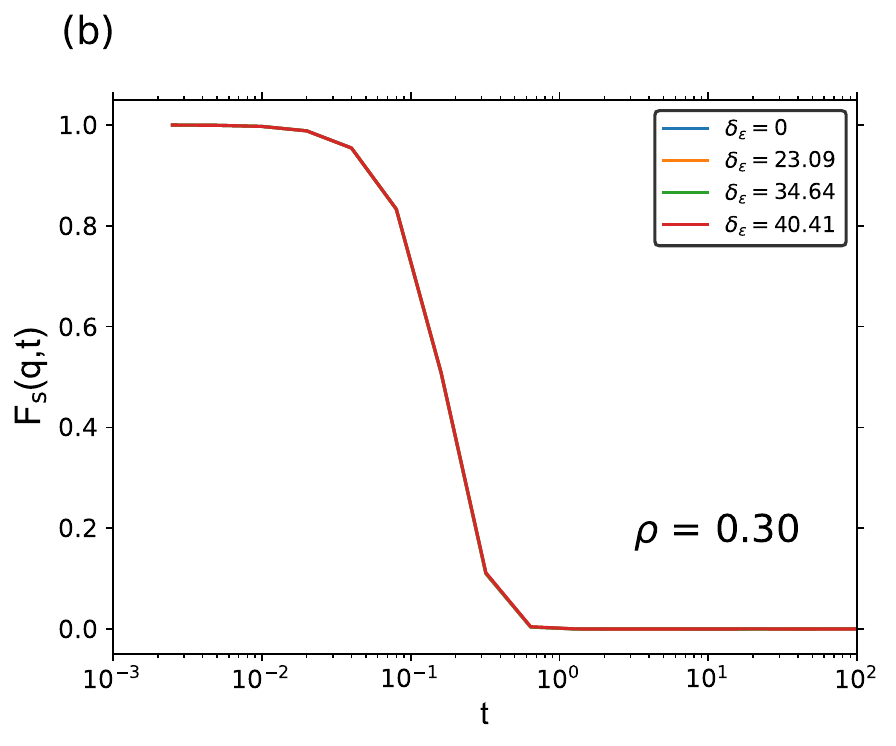}
  \end{minipage}
    \centering
  \begin{minipage}[t]{0.45\linewidth}
  \centering
    \includegraphics[width=70mm]{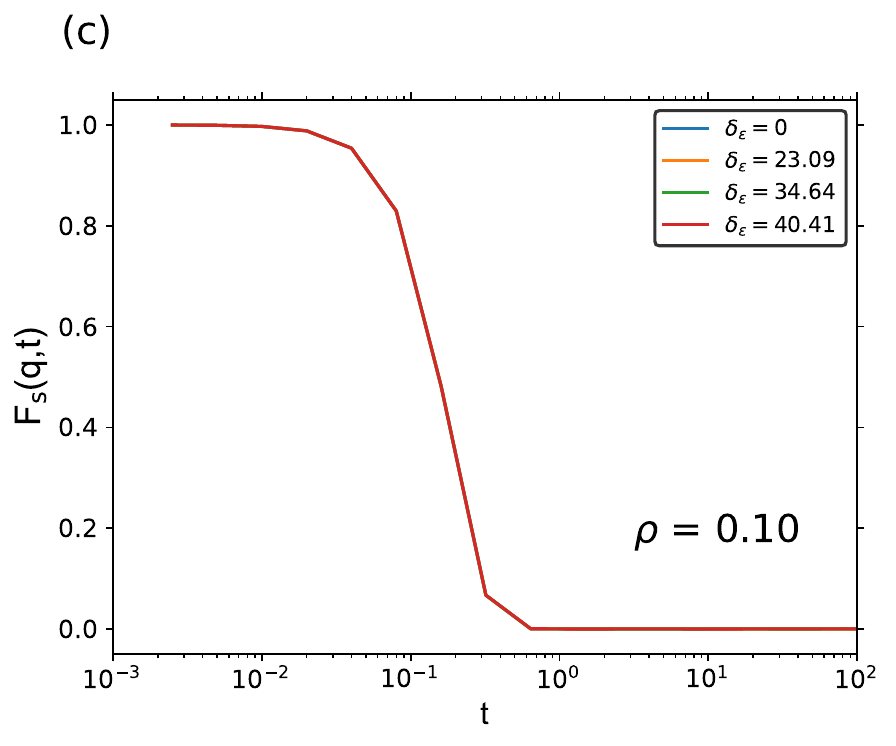}
    \end{minipage}
  \caption{Effect on the average ISF of different energy polydispersities at $T$ = 1.3 at three different densities. (a) $\rho$ = 0.50. (b) $\rho$ = 0.30. (c) $\rho$ = 0.10.}
  \label{lowerdensityfs}
\end{figure}

We plot in Fig. \ref{lowerdensityfs} the average ISF at the same state points. The same conclusion is reached as for the structure: energy polydispersity leads to little change from the results of the monodisperse system.

All results reported in this subsection are for $T=1.3$. There is nothing special about this temperature, however; the average structure and dynamics is almost independent of the degree of energy polydispersity also at $T$ = 2 (results not shown).

\subsection{Dependence on the energy mixing rule}

\begin{figure}[H]
    \centering
    \includegraphics[width=70mm]{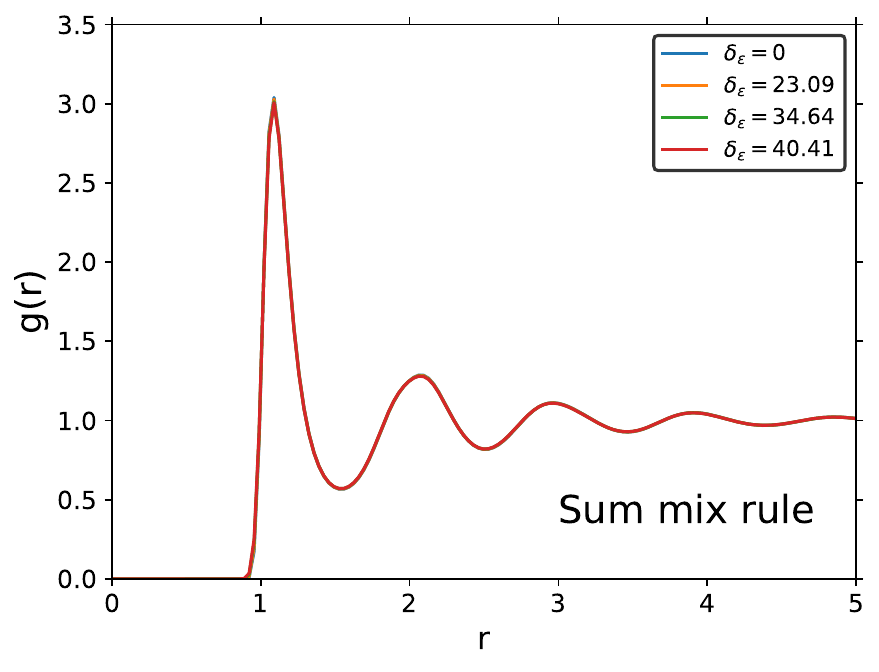}
  \caption{The effect of energy polydispersity at  $(\rho,T)= (0.85,0.70)$ using the ``sum mix rule'' arithmetic mean of the particle energies, Eq. (\ref{new}), for defining the interaction energy $\varepsilon_{ij}$.}
  \label{arit}
\end{figure}

What happens if one changes the energy mixing rule to the arithmetic instead of the geometric mean of the Lorentz-Berthelot mixing rule? Replacing Eq. (\ref{mix}) by

\begin{align}
    \varepsilon_{ij} = \frac{\varepsilon_{i} + \varepsilon_{j}}{2}\,.
    \label{new}
\end{align}
leads to the results of Fig. \ref{arit} for the average RDF.
We see that, even for this significant change of the energy mixing rule, the average structure is still almost independent of the energy polydispersity. The same applies for the dynamics (results not shown).

\subsection{Using a Gaussian instead of a box energy distribution}

\begin{figure}[H]
    \centering
    \includegraphics[width=70mm]{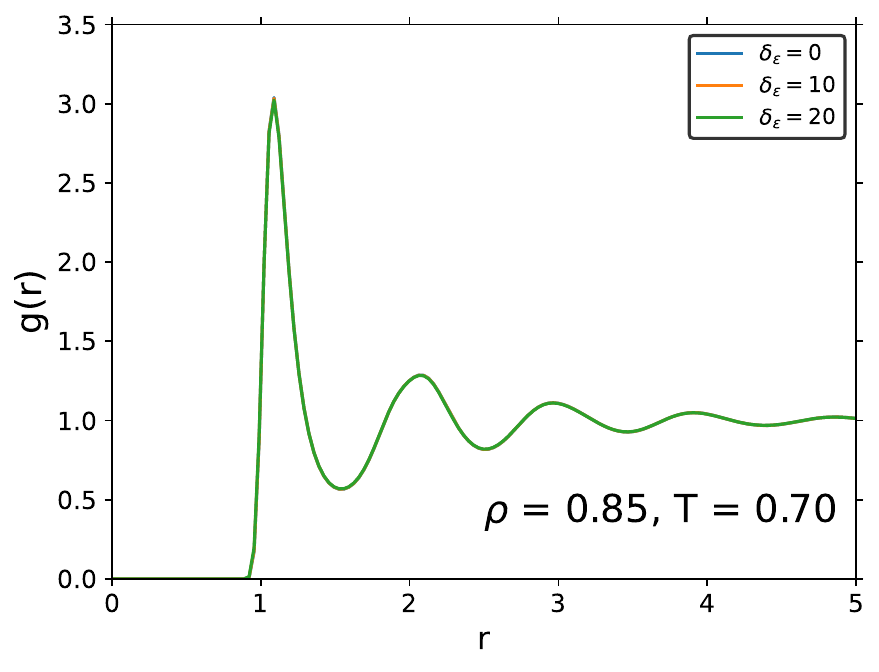}
  \caption{Effect of energy polydispersity at  $(\rho,T)= (0.85,0.70)$ using a Gaussian distribution.}
  \label{gauss}
\end{figure}

In order to investigate whether there is something special about the box distribution, we studied the effect of using instead a Gaussian distribution. In this case we had to limit ourselves to lower polydispersities due to the long tail of the Gaussian distribution. The results for the average RDF (Fig. \ref{gauss}) and the average ISF (not shown) are entirely analogous to those of the box distribution, however.

\subsection{The exponential pair potential}

\begin{figure}[H]
  \begin{minipage}[t]{0.45\linewidth}
    \centering
    \includegraphics[width=70mm]{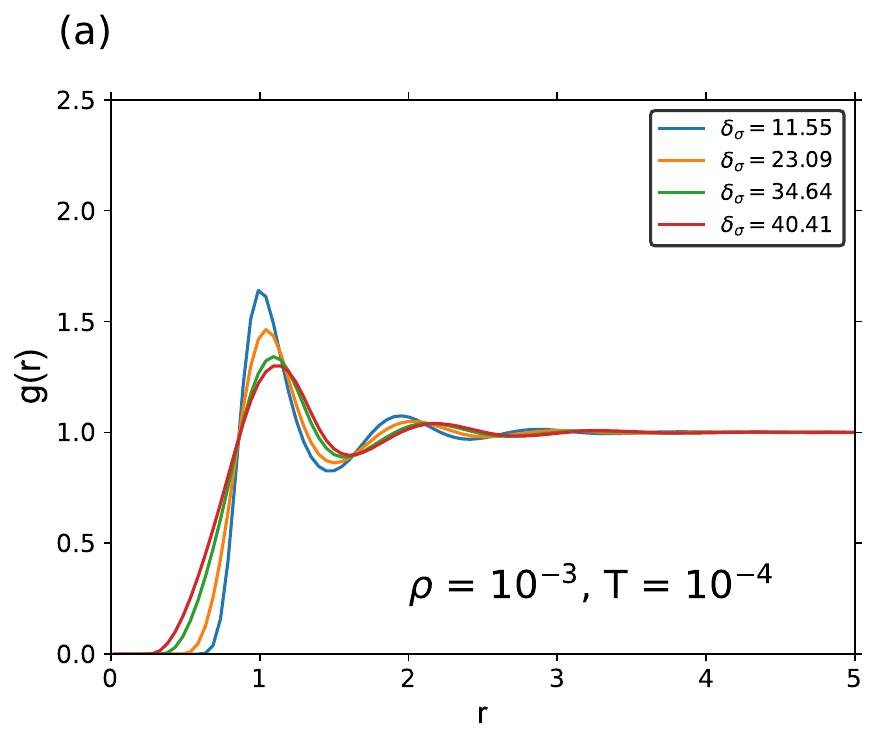}
  \end{minipage}
  \hspace{0.1cm}
  \begin{minipage}[t]{0.45\linewidth}
    \centering
    \includegraphics[width=70mm]{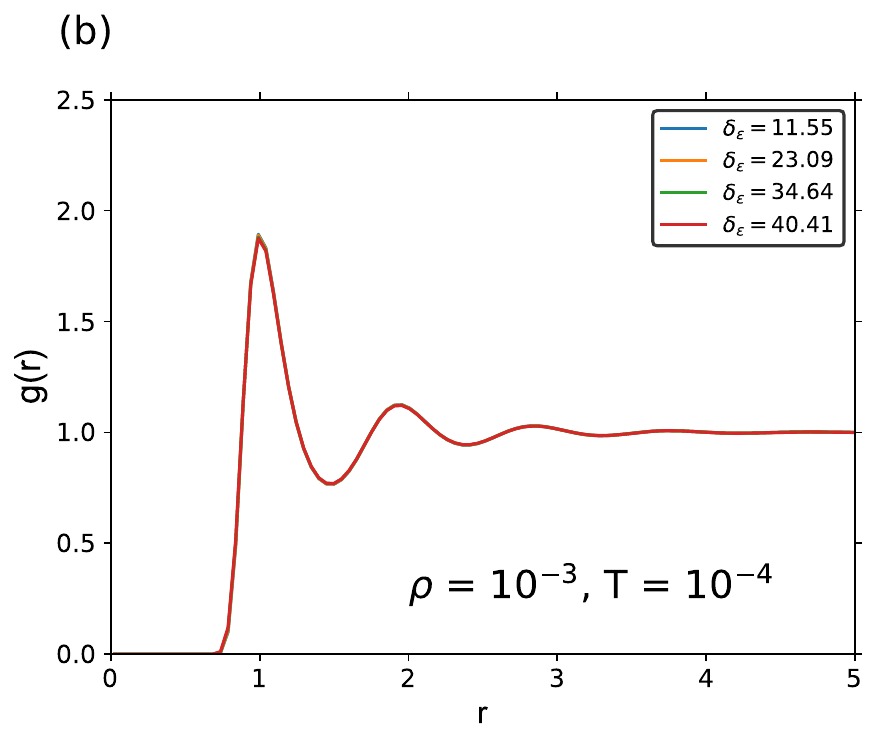}
  \end{minipage}
 \begin{minipage}[t]{0.45\linewidth}
    \centering
    \includegraphics[width=70mm]{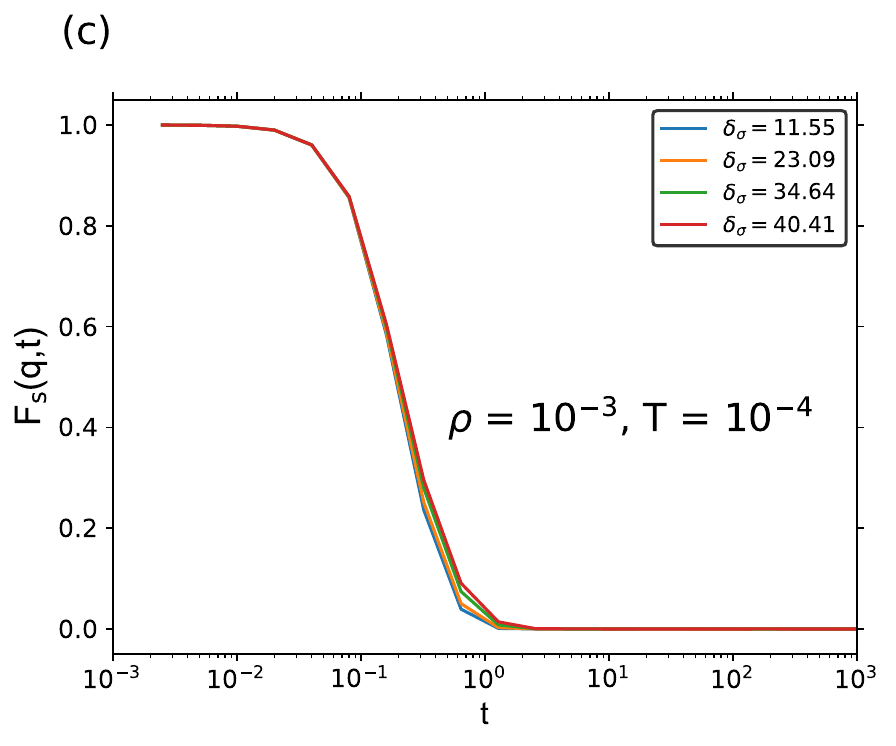}
  \end{minipage}
  \hspace{0.5cm}
  \begin{minipage}[t]{0.45\linewidth}
    \centering
    \includegraphics[width=70mm]{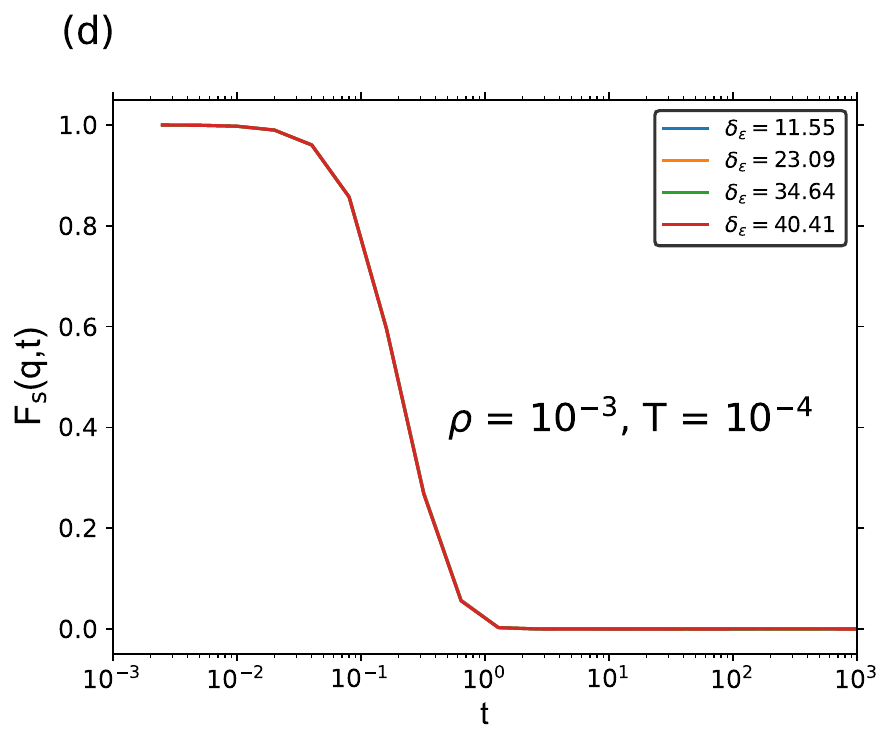}
  \end{minipage}
   \caption{Effect of introducing size and energy polydispersity for the EXP pair potential at the state point $(\rho, T) = (10^{-3},10^{-4})$. (a) and (c) give results for size polydispersity, while (b) and (d) give results for energy polydispersity.}
  \label{exp}
\end{figure}

Next we address whether the insensitivity to the degree of energy polydispersity is particular to the LJ pair potential. A quite different pair potential is the exponential repulsive EXP pair potential, which has been argued to be the ``mother of all pair potentials'' in the sense that the quasiuniversality of simple liquids may be explained in terms of it \cite{bacher2014n,dyr16}. The EXP pair potential is defined by

\begin{align}
v_{ij}(r) = \varepsilon_{ij} \exp(-r/\sigma_{ij})\,.
\end{align}
For studying the effect of energy polydispersity on the EXP pair potential we return to using the box distribution and the Lorentz-Berthelot mixing rules. The results for the average structure and dynamics for size (left) and energy (right) polydispersity are given in Fig. \ref{exp}

As previously, very similar average structure and dynamics is observed with varying degree of energy polydispersity, while strong effects are observed for varying size polydispersity.

\subsection{Results for binary LJ mixtures}

\begin{figure}[H]
  \begin{minipage}[t]{0.45\linewidth}
    \centering
    \includegraphics[width=70mm]{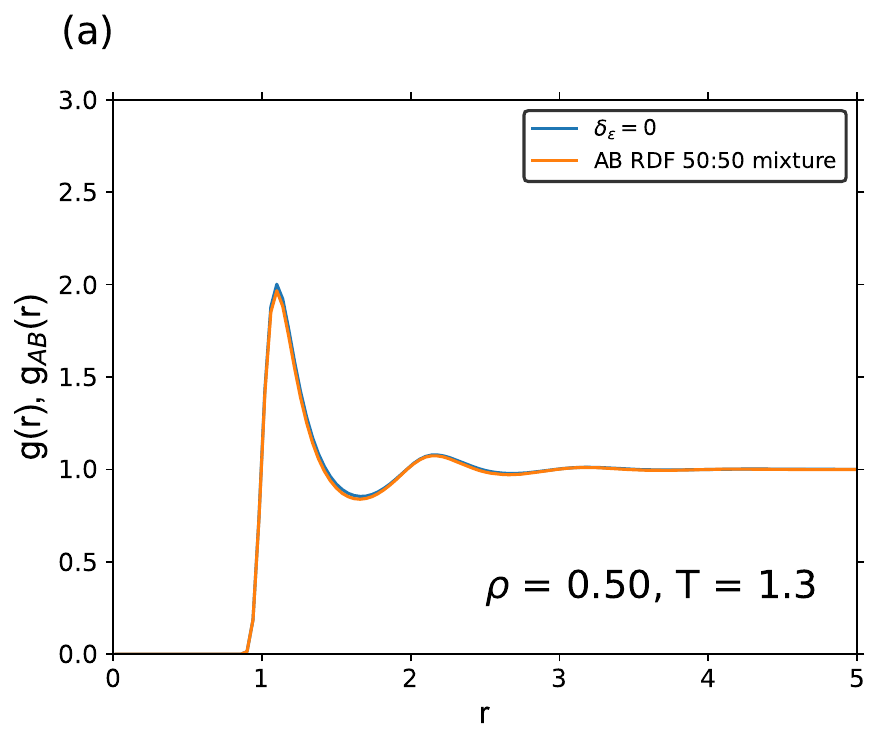}
  \end{minipage}
  \hspace{0.5cm}
  \begin{minipage}[t]{0.45\linewidth}
    \centering
    \includegraphics[width=70mm]{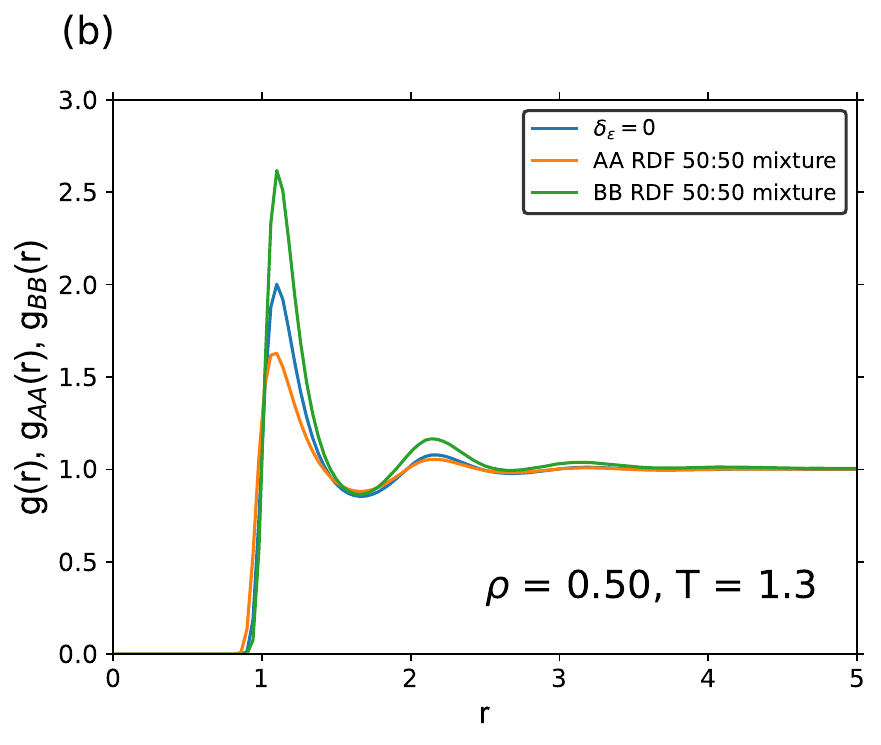}
  \end{minipage}
  \caption{The effect of energy polydispersity on the structure at $(\rho,T)= (0.5,1.350)$ for a 50:50 binary mixture with $\epsilon_{AA} = 1.5$, $\epsilon_{BB} = 0.5$, $\epsilon_{AB} = 1$. g$_{AB}$(r) is plotted in (a), while results for the $AA$ and $BB$ RDFs are shown in (b).}
  \label{binary}
\end{figure}

An alternative to the LJ energy polydispersity studied in Sec. III is to replace continuous polydispersity by a 50:50 AB binary mixtures with large energy variation. This deviates from the present paper's focus on continuous distributions, but is nevertheless worth attention. Figure \ref{binary} shows results for a mixture with a factor of three different particle energies. In this case we observe a virtually unchanged mean $g_{AB}(r)$ (Fig. \ref{binary}(a)), while the average $g_{AA}(r)$ or $g_{BB}(r)$ \textit{are} affected by the energy polydispersity ((b)).

\section{Phase separation}

\begin{figure}[H]
  \centering
  \includegraphics[width=80mm]{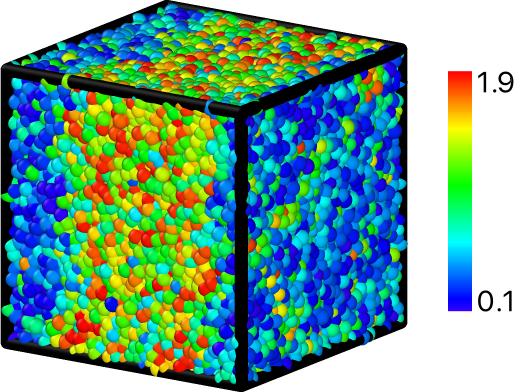}
  \caption{Snapshot of a configuration at $\delta_{\varepsilon}$ = 52\%, corresponding to almost a factor of 20 energy variation. A strong but continuous segregation is observed.}
  \label{snap}
\end{figure}

Previous studies have shown that energy-polydisperse systems phase separate into lower and higher particle-energy regions \cite{shagolsem2015a,ingebrigtsen2016}. An example of this is shown in Fig. \ref{snap} for a very high energy polydispersity (52\%). This behavior is not an artifact of the thermostat since using $NVE$ and Langevin dynamics leads to the same behavior (results not shown).

\begin{figure}[H]
  \begin{minipage}[t]{0.45\linewidth}
    \centering
    \includegraphics[width=70mm]{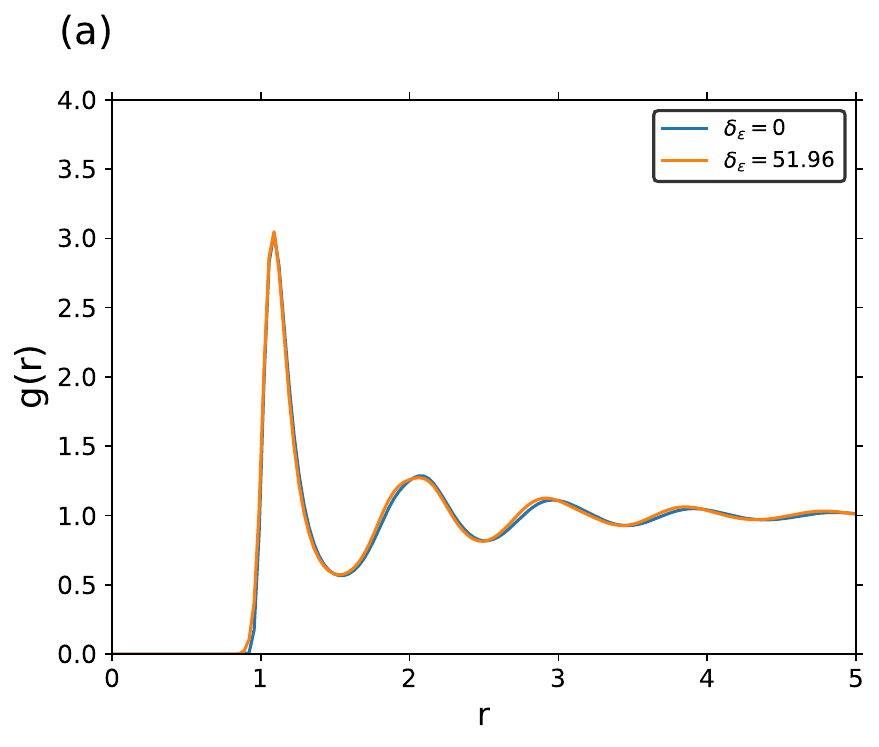}
  \end{minipage}
  \hspace{0.5cm}
  \begin{minipage}[t]{0.45\linewidth}
    \centering
    \includegraphics[width=70mm]{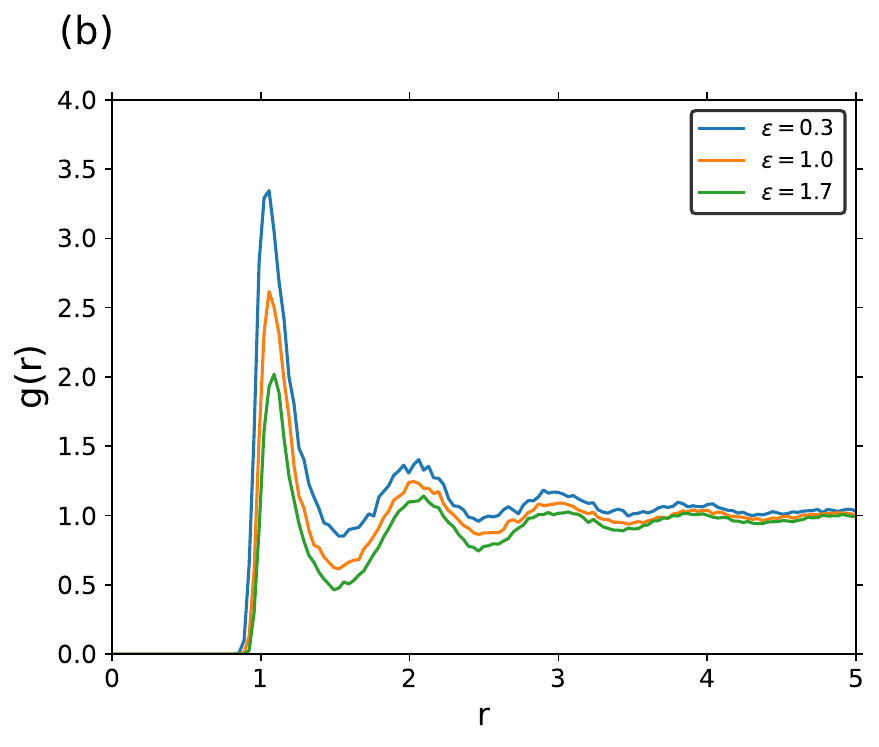}
  \end{minipage}
      \caption{(a) Comparing the average RDFs for zero and a very high energy polydispersity (52\%). (b) Energy-resolved average RDFs. The local structure around each particle correlates with its energy. This figure shows data for energy polydispersity $\delta_{\varepsilon}$ = 40.41\% with the particles binned with $\Delta \varepsilon$ = 0.1.}
  \label{lowhigh}
\end{figure}

In the extreme case $\delta_{\varepsilon}$ = 52\% of Fig. \ref{snap} we do find visible changes in the average RDF, albeit very minor ones (Fig. \ref{lowhigh}(a)). A possible explanation of these changes could be that all the energy polydisperse potentials share the same harsh repulsion and can be mapped onto a hard-sphere system with a similar effective radius. If so, one would expect not only the average structure, but also the \emph{local} structure around each particle to be unaffected by energy polydispersity. Figure \ref{lowhigh}(b) shows that this is not the case, however; the RDF of a given particle clearly correlates with its energy $\varepsilon_i$. This shows that one cannot explain our results by appealing to an equivalent hard-sphere system.

\begin{figure}[H]
    \centering
    
\end{figure}

\section{Discussion}

Polydisperse liquids are intriguing systems with a rich phenomenology. Size polydispersity is by far the most commonly studied type of polydispersity, but lately energy polydisperse systems have also gained interest. In this paper we have compared these two kinds of polydispersity in regard to how the average structure and dynamics are affected. We found that size polydispersity strongly influences both the average structure and the average dynamics, whereas energy polydispersity -- even for quite strong polydispersity -- has only an insignificant effect. The local structure is, however, affected also in the latter case for which higher-energy particles tend to cluster and separate spatially from the lower-energy particles. It should be pointed out that the phase separation eventually leads to crystallization in very long simulations; all results reported above refer to the average structure and dynamics before there are signs of crystallization.

We consider the findings for energy polydisperse systems to be striking and do not have a good explanation. On a qualitative level the independence of the physics on the degree of energy polydispersity is consistent with the main physical assumptions of the old perturbation theories of Weeks-Chandler-Andersen \cite{wca} and Barker-Henderson \cite{bar76}. These theories featured dominance of the short range excluded-volume interactions in determining the microstructure of a liquid, and it is reasonable to assume that the excluded-volume interactions are only weakly affected by the strength of the repulsive interactions, i.e., by the energy polydispersity. This ``explanation'' is, however, challenged by the results reported in Fig. \ref{lowhigh}(b).

One of our results conforms to the predictions of the van der Waals mixing rule of conformal solution theory \cite{man93,she06}. The idea of the ``one-fluid approximation''  of conformal solution theory is that the mixture in question can be represented as a single-component fluid. For the case of size polydispersity, it is well known that serious problems arise when characterizing the average RDF of even moderately polydisperse fluids by an effective one-component approach \cite{pond2011}. Various approximations have been proposed for defining the one-fluid approximation to a given polydisperse system \cite{man93}. One of the simplest is the so-called van der Waals mixing rule according to which the energy parameter of the ``one-fluid'' representing the mixture is $\langle\varepsilon_{ij}\rangle$. For the above-studied cases of a symmetrical energy distribution, assuming the linear energy mixing rule Eq. (\ref{new}) is equivalent to assuming the van der Waals mixing rule. 

It is our hope that this paper inspires to the development of a theoretical framework explaining the insensitivity of the physics to the introduction of energy polydispersity. A striking finding usually has a simple explanation, and there seems to be no reason this should not apply also for energy-polydisperse liquids.

\section*{Acknowledgement}

This project was initiated under a Japan Society for the Promotion of Science (JSPS) postdoctoral fellowship (T.S.I.). The work was subsequently supported by the VILLUM Foundation’s \textit{Matter} grant (No. 16515). We are grateful to Hajime Tanaka, Daniele Dini, David Heyes, Ulf Pedersen, and Thomas Schr{\o}der for stimulating discussions.

\end{document}